# Self-Adaptive 2D-3D Ensemble of Fully Convolutional Networks for Medical Image Segmentation

Maria G. Baldeon Calisto, Susana K. Lai-Yuen
University of South Florida
mariabaldeon@mail.usf.edu, laiyuen@usf.edu

## ABSTRACT

Segmentation is a critical step in medical image analysis. Fully Convolutional Networks (FCNs) have emerged as powerful segmentation models achieving state-of-the art results in various medical image datasets. Network architectures are usually designed manually for a specific segmentation task so applying them to other medical datasets requires extensive expertise and time. Moreover, the segmentation requires handling large volumetric data that results in big and complex architectures. Recently, methods that automatically design neural networks for medical image segmentation have been presented; however, most approaches either do not fully consider volumetric information or do not optimize the size of the network. In this paper, we propose a novel self-adaptive 2D-3D ensemble of FCNs for medical image segmentation that incorporates volumetric information and optimizes both the model's performance and size. The model is composed of an ensemble of a 2D FCN that extracts intra-slice information, and a 3D FCN that exploits inter-slice information. The architectures of the 2D and 3D FCNs are automatically adapted to a medical image dataset using a multiobjective evolutionary based algorithm that minimizes both the segmentation error and number of parameters in the network. The proposed 2D-3D FCN ensemble was tested on the task of prostate segmentation on the image dataset from the PROMISE12 Grand Challenge. The resulting network is ranked in the top 10 submissions, surpassing the performance of other automatically-designed architectures while being considerably smaller in size.

## 1. INTRODUCTION

Accurate segmentation is crucial to various medical tasks such as studying anatomical structures, measuring tissue volume, and assisting in treatment planning before radiation therapy [1]. Fully convolutional networks (FCNs) have been shown to provide state-of-the-art results for medical image segmentation. However, FCN architectures are normally designed manually for a specific medical segmentation task. Given the high complexity and depth of current architectures, manually adapting the architectures to a new dataset resembles a black-box optimization process that requires extensive experience, time, and computational resources.

Two main types of FCNs have been proposed for handling volumetric medical image data. The first models are 2D networks that segment images in 2D and then concatenate them to provide the 3D segmentation [2, 3, 4]. Although these methods are able to capture rich information in one plane, they do not fully exploit the spatial correlation along the z-axis. The second type of networks are 3D FCNs that replace 2D convolutions with 3D convolutions and directly process volumetric information [5, 6, 7]. Nevertheless, 3D FCNs need a substantial number of parameters to capture representative features, and require high computational time and GPU memory. Recent work has focused on hybrid 2D-3D FCNs to combine the strengths of 2D and 3D FCNs [8, 9]. However, these 2D-3D architectures remain considerably big and comparable in size and memory consumption to other 3D FCNs.

To address these challenges, there has been an increasing focus on developing methods that automatically design neural networks through the application of optimization algorithms, also known as neural architecture search (NAS). NAS can be considered a subfield of auto machine learning (AutoML) and has a significant overlap with hyperparameter optimization and meta-learning [10]. The algorithms applied in NAS have used reinforcement learning [11, 12, 13], evolutionary algorithms [14, 15, 16], surrogate model-based optimization [17], and one-shot architecture search [18, 19].

However, recent neural architecture search (NAS) methods have been proposed mainly for image classification and language modeling [10, 20].

For medical image segmentation, the use of NAS has been limited. Isensee et al. [21] presented a self-adapting framework that uses a rule-based approach to determine the pre-processing operations and training parameters of a pool of U-Net architectures (2D U-Net, 3D U-Net and cascaded U-Net). In [22], Mortazi and Bagci proposed a policy gradient reinforcement learning based method to find the hyperparameters of a 2D densely connected encoder-decoder baseline CNN. In [23], Weng et al. proposed three types of primitive operation sets to construct down-sampling and up-sampling cells for a 2D U-Net backbone network where the cell configurations are updated using the DARTS [18] differential search strategy. In our previous work [24], we proposed an adaptive 2D U-Net inspired architecture called AdaResU-Net, which applies a multiobjective evolutionary based algorithm to search for the hyperparameters of a semi-fixed architecture. This resulted in adaptive models that maximize the segmentation accuracy and minimize the model's size for 2D segmentation. In [25], Zhu et al. presented a differentiable NAS that selects between 2D, 3D or Pseudo-3D convolutions for each layer of a FCN architecture. Similarly, Kim et al. [26] proposed a 3D U-Net template architecture that finds the configuration for the encoder, reduction, decoder and expansion cells by applying a differentiable NAS algorithm. Previous approaches either design 2D networks that segment the 3D images in a slice-wise manner, which does not consider crucial volumetric information, or find 3D configurations that rely on predefined architecture templates of fixed depth that do not optimize the size of the network.

In this paper, we present a self-adaptive 2D-3D FCN ensemble for medical image segmentation that incorporates volumetric information and optimizes both the model's performance and its size. The network is composed of a 2D FCN that extracts in-plane information and a 3D FCN that exploits volumetric information. Both FCN architectures are automatically fitted to a specific medical image dataset using a multiobjective evolutionary based algorithm that maximizes segmentation accuracy and minimizes the number of parameters in the network. In contrast to other methods for medical image segmentation, our model is self-adaptive by searching for the optimal hyperparameters and architecture, fully utilizing volumetric information, and minimizing the size of the network. The proposed 2D-3D FCN ensemble was tested in the task of prostate segmentation on the PROMISE12 Grand Challenge [27]. Our model is ranked within the top 10 submissions of the leaderboard surpassing the performance of automatically-designed architectures while being considerably smaller in size. Therefore, we demonstrate that the proposed model can successfully adapt the architecture configuration and hyperparameters to the dataset. We also show the potential of multiobjective evolutionary based algorithms for automatically designing smaller and efficient neural networks for medical image segmentation.

## 2. METHODS

The proposed 2D-3D FCN ensemble is constructed in two phases as shown in Fig. 1. In Phase I, the 2D FCN and 3D FCN architectures are adapted to the specific dataset using a Multiobjective Evolutionary based Algorithm (MEA algorithm) presented in our previous work [24]. This is performed by dividing the dataset into 5 folds and selecting a fold at random to define the 2D and 3D FCN architectures. In Phase II, the optimal 2D FCN and 3D FCN architectures are trained with each of the 5 folds from the training dataset. For each fold, the predictions of the 2D FCN and 3D FCN are averaged and the final prediction is obtained by a majority voting of the five 2D-3D ensemble models.

Phase I begins by dividing the dataset into 5-folds. One fold is selected at random to define the architecture of the 2D FCN and 3D FCN, where the training set images are used to train the candidate FCNs and the validation set images to evaluate the performance of the architecture. A view of a five residual-block FCN architecture is shown in Fig. 2. Each FCN has a symmetrical encoder-decoder structure where the basic building module is a residual block. A residual block is composed of three convolutional stages, where each stage is comprised of a convolutional layer, batch normalization layer [28], and an activation function layer. The first and last convolutional stages in a residual block are connected through a residual connection [29] to improve gradient flow and information transmission. The proposed FCN has the same number of residual blocks on the encoder and decoder structure. Furthermore, a merge operation is implemented between residual blocks located on opposite sides of the encoder-decoder structure to enable low-level feature preservation [3]. In the encoder structure, the residual block is followed by a max-pooling operation of stride 2 that halves the size of the input feature maps. On the other hand, the decoder structure is implemented by a transpose convolution that doubles the size of the input feature maps. Finally, to prevent overfitting, a spatial dropout layer is located before a residual block [30]. The architecture for the 2D FCN and 3D FCN is the same with the exception that the 2D FCN applies 2D convolutions while the 3D FCN uses volumetric convolutions. The last convolutional layer of the architecture has a fixed kernel of size 1 and a softmax activation function.

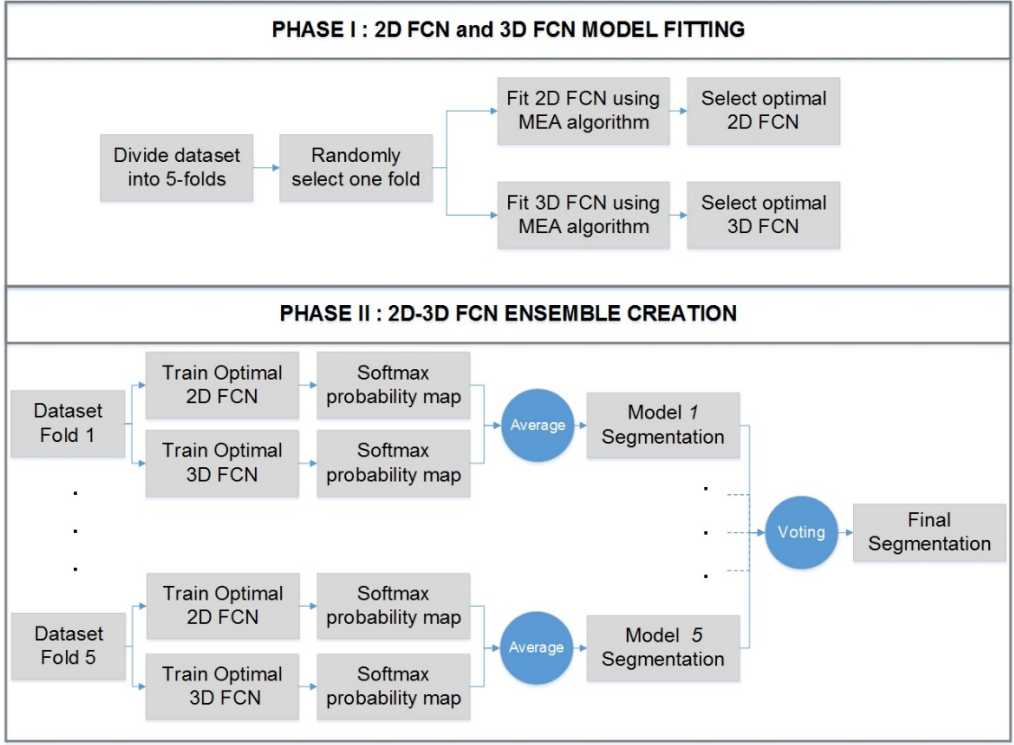

Figure 1. Overview of the proposed 2D-3D FCN ensemble construction. In Phase I, one randomly selected fold is used to adapt the 2D FCN and 3D FCN. In Phase II, the 5 folds are used to train the architectures found in Phase I, creating the 2D-3D FCN ensemble network.

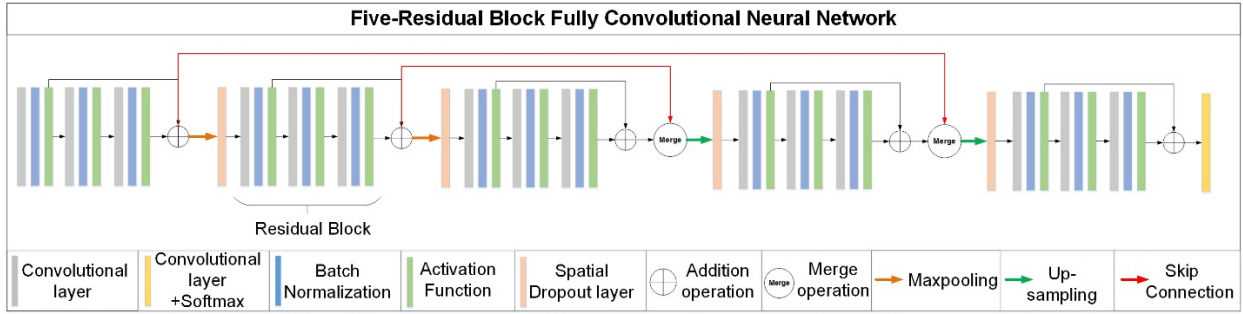

Figure 2. Example of a five-residual block FCN.

Once the overall structure of the FCN is defined, nine hyperparameters that are encoded into nine decision variables need to be set to construct the final architecture. As shown in Table 1, these hyperparameters are: (1) total number of residual blocks, (2) number of filters on each residual block, (3-5) kernel size of each convolutional layer on a residual block, (6) activation function, (7) merge operation between encoder and decoder residual blocks, (8) spatial dropout probability, and (9) learning rate.

For the first decision variable, the total number of residual blocks ($N_{block}$) defines the depth of the architecture. A number of $\frac{N_{block}}{2} - 0.5$ residual blocks are located in the encoder structure as well as in the decoder structure while one residual block connects the encoder-decoder structure. The number of filters in each residual block are computed using the following rule: the number of filters is doubled after a max-pooling operation and halved after a transpose convolution operation. Thus, the number of filters in the entire architecture can be calculated based on the number of filters assigned to the first residual block (second decision variable) and using the stated rule. Three variables encode the kernels size of the three convolutional layers inside the residual block. We apply square or cubic (in case of 3D convolutions) kernels so only one variable is needed to define the kernel size of a convolutional layer. Another three decision variables are used to define the activation function, merge operation, and spatial dropout probability that are applied through the entire architecture. Finally, one decision variable is used to define the learning rate. The nine decision variables and their corresponding search space are presented in Table 1.

Table 1. Set of hyperparameters and their corresponding search space for the construction of the 2D FCN and 3D FCN.

| Hyperparameter | 2D FCN Search Space | 3D FCN Search Space |
| --- | --- | --- |
| Residual Blocks | [3,5,7,9] | [3,5,7,9] |
| Number of filters in the first residual block | [4,8,16,32] | [4,8,16,32] |
| Kernel size for Conv. layer 1 | [1×1, 3×3, 5×5, 7×7] | [1×1×1, 3×3×3, 5×5×5] |
| Kernel size for Conv. layer 2 | [1×1, 3×3, 5×5, 7×7] | [1×1×1, 3×3×3, 5×5×5] |
| Kernel size for Conv. layer 3 | [1×1, 3×3, 5×5, 7×7] | [1×1×1, 3×3×3, 5×5×5] |
| Activation Function | [ReLu, elu] | [ReLu, elu] |
| Merge Operation | [Summation, Concatenation] | [Summation, Concatenation] |
| Dropout Probability | [0, 0.7] | [0, 0.7] |
| Learning Rate | [$1\times10^{-8}$, $9\times10^{-3}$] | [$1\times10^{-8}$, $9\times10^{-3}$] |

The hyperparameters of the final architecture are learned using our MEA algorithm [24], which is a population based optimization method that utilizes the MOEA/D algorithm [31] and a Penalty-based Boundary Intersection approach to approximate the non-dominated Pareto Frontier. The MEA algorithm defines the architecture by minimizing two objective functions: the expected segmentation error and the number of parameters. The expected segmentation error is quantified through the Dice similarity coefficient (*DSC*) that measures the overlap between the predicted segmentation and the ground truth segmentation and is defined as:

$$DSC = \frac{2\sum_i \hat{y}_i y_i}{\sum_i \hat{y}_i + \sum_i y_i} \quad (1)$$

Where $\hat{y}_i$ are the voxels from the predicted segmentation and $y_i$ the voxels from the ground truth segmentation. The expected segmentation error is quantified through the *DSC* in the training set, the *DSC* in the validation set and a term that considers the epoch with the maximum validation *DSC*. In this work, we use partial training to reduce the convergence time to an optimal FCN. Thus, the last term accounts for the expected improvement if the candidate FCN is trained for additional epochs. The optimization problem the MEA algorithm solves is:

$$Minimize\ f_1(x) = \alpha\big(1 - DSC_{Train}(\theta)\big) + \big(1 - DSC_{Val}(\theta)\big) + \beta\left(\frac{E - e_{max}}{E}\right) \quad (2)$$
$$Minimize\ f_2(x) = |\theta| \quad (3)$$
$$subject\ to\ x \in \Omega \quad (4)$$

Where $DSC_{Train}(\theta)$ and $DSC_{Val}(\theta)$ are the Dice similarity coefficient in the train and validation set, respectively. $E$ is the maximum number of epochs the candidate FCN is trained, $e_{max}$ is the epoch number in which the maximum validation *DSC* is obtained, $\alpha$ and $\beta$ are weight parameters whose value is between the range [0, 1], $\theta$ are the parameters in the FCN, and $|\cdot|$ is the cardinality operator. Finally, $x$ is the hyperparameter vector with the nine components shown in Table 1 that define the FCN architecture, and $\Omega$ the hyperparameter search space. The weight parameters balance the importance of the *DSC* in the training set and expected segmentation improvement over the *DSC* on the validation set while searching for the optimal architecture. Since obtaining an architecture that is capable of generalizing well to unseen data is crucial, setting $\alpha$ and $\beta$ to a small value is recommended. After experimentation, we have found $\alpha = 0.25$ and $\beta = 0.25$ to be appropriate.

The MEA algorithm is initialized by randomly selecting hyperparameter values for the first population of architectures. These architectures are trained through backpropagation algorithm and their fitness quantified through objective functions (2) and (3). In each generation afterwards, candidate architectures are generated by applying cross-over and mutation operations to the current solutions. These architectures are trained through backpropagation and their segmentation error and number of parameters computed. The current solution is updated using a Penalty-based Boundary Intersection approach. The algorithm terminates when a specified number of generations has elapsed. From the returned Pareto non-dominated Frontier, the hyperparameters that minimize loss function (2) are selected for the optimal architecture.

After finding the optimal 2D and 3D FCN architectures, Phase II constructs the 2D-3D ensemble. This is accomplished by training the 2D FCN and 3D FCN optimal architectures with each of the 5-folds from the training dataset and subsequently averaging the softmax probability maps of the 2D and 3D FCNs. Thus, five 2D-3D ensemble models are trained resulting in five predicted segmentations. The final image segmentation is determined by using a majority voting.

Ensemble networks have shown to yield a comparable or higher accuracy than the best individual network if the members of the ensemble are accurate and make less correlated errors [32]. In this work, we attain diversity by constructing a different type of FCN for each training fold (2D vs. 3D structure), initializing the weights in the FCN with random values, and training each 2D-3D FCN with distinct folds.

## 3. RESULTS

### 3.1 Implementation Results

The 2D-3D FCN ensemble is constructed using the dataset from the MICCAI 2012 Prostate MR Image Segmentation challenge [27]. The dataset is comprised of 50 transversal T2-weighted MR images from the prostate with their corresponding ground truth for training and 30 MR images without reference for testing. Since the images have a high variance in size, resolution and appearance, we resample all to a spatial resolution of 1×1×1.5 mm and set to a fixed size of 128×128×64 voxels. Also, pixel intensities are clipped to be inside the 3 standard deviations from the mean and rescaled to a 0-1 range. For the 2D FCN, the input are slices of size 128×128 whereas the 3D FCN is trained with randomly extracted patches of size 96×96×16. The experiments are carried on an 8-GB NVIDIA GeForce GTX 1080 GPU, 3.60-GHz CPU and 16-GB RAM.

The MEA algorithm evolves for 40 generations, using a penalty factor of 0.1 and training a maximum of 120 epochs per candidate architecture. The other parameters are set as described in [24], with the exception of the mutation probability that is applied independently to each component in the hyperparameter vector with a value that decreases monotonically. This decrease allows the search algorithm to explore distinct areas in the initial generations whereas allowing convergence towards the end of the evolution. In Table 2 the optimal hyperparameter values found for the 2D FCN and 3D FCN are shown as well as the number of parameters for each architecture. The 2D FCN search took approximately 66.31 hours and the 3D FCN search 118.08 hours.

Table 2. Optimal hyperparameters and number of parameters of the optimal 2D FCN and 3D FCN.

| Hyperparameter | 2D FCN | 3D FCN |
|---|---|---|
| Residual Blocks | 7 | 5 |
| Number of Filters | 16 | 32 |
| Kernel Size Conv. 1 | 1×1 | 3×3×3 |
| Kernel Size Conv. 2 | 3×3 | 1×1×1 |
| Kernel Size Conv. 3 | 7×7 | 5×5×5 |
| Activation Function | ReLU | elu |
| Merge Operation | Concatenation | Concatenation |
| Dropout Probability | 0.15 | 0 |
| Learning Rate | $4\times10^{-4}$ | $5\times10^{-5}$ |
| **Number of Parameters** | $1.6 \times10^{6}$ | $3.9 \times10^{6}$ |

Once the optimal hyperparameters were obtained, five 2D-3D ensembles are trained with the 5-fold data division and using the ADAM optimizer [33]. Data augmentation is applied to enlarge the training dataset by including random rotation, scaling, vertical and horizontal translation and horizontal flip of the original images. The model is implemented with Python 3.6 and using Keras library [34]. The 2D FCNs are trained for 3000 epochs and the 3D FCN for 6000 epochs. The weights with the minimum validation loss are used for predicting the final segmentations. To yield the test segmentation results, the final segmentations produced by the ensemble are resampled to the original image resolution. Also, a connected component analysis is applied as post-processing operation where only the largest component is kept. Fig. 3 shows the qualitative results of our model on the validation set. The resulting segmentations have a spatially consistent shape and smooth boundaries, even though no shape prior or task-specific mechanisms were introduced during training or in the architecture.

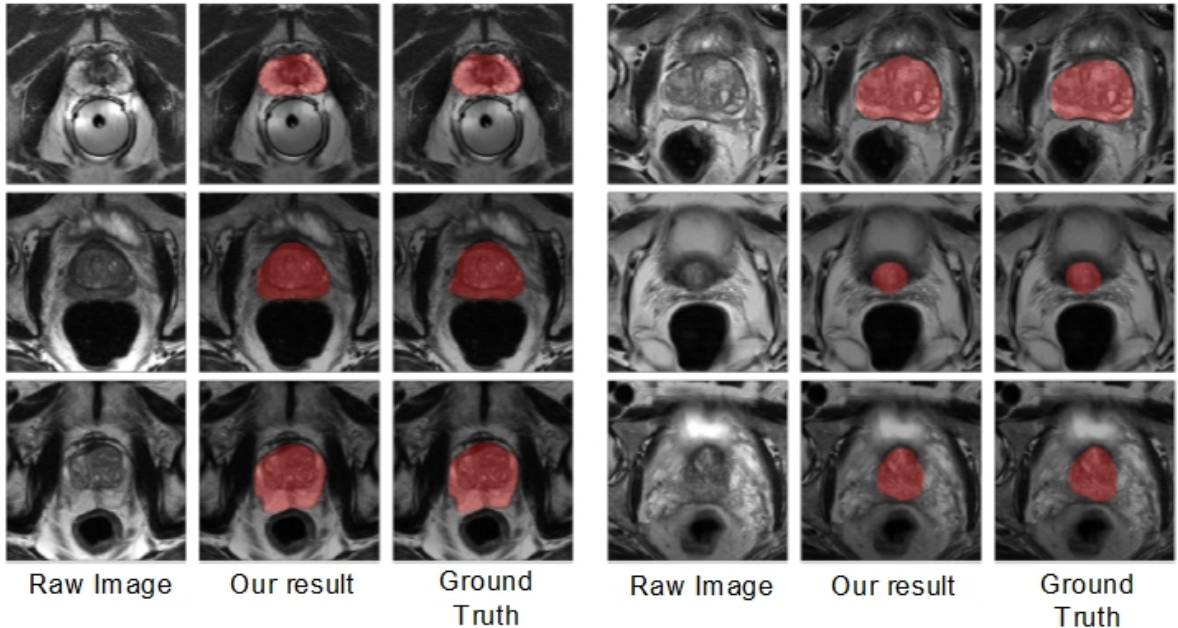

Figure 3. Example of our segmentation results on the PROMISE12 dataset. The network produces spatially consistent segmentation with smooth boundaries.

## 3.2 Comparison with State-of-the-art

The evaluation of the test cases were carried via an online submission to the PROMISE12 challenge. Four evaluation metrics are used to assess the volumetric segmentations of the whole prostate, apex and base parts of the prostate. The evaluation metrics include the Dice Similarity Coefficient (DSC), 95 percentile Hausdorff distance (95 HD), average boundary distance (ABD), and absolute relative volume difference (aRVD). Also, the challenge ranks each method by computing an overall score that combines all the evaluation metrics for the 30 test cases. Table 3 presents the ranking of the proposed method (2D-3D FCN) and top competing models on the leaderboard and Table 4 the evaluation metrics. Only methods that have submitted a description of their methods are shown for a fair comparison.

Table 3. Ranking of the proposed method (2D-3D FCN) and top competing models on the PROMISE12 challenge test set.

| Model | Design Type | Overall Score | Rank |
|---|---|---|---|
| 2D-3D FCN (ours) | Automatic | 89.293 | 9 |
| HD_Net | Manual | **90.344** | **1** |
| Bowda-Net | Manual | 89.585 | 2 |
| Sunrise2014 | Manual | 89.461 | 6 |
| nnU-Net (I) | Automatic | 89.276 | 10 |
| nnU-Net (II) | Automatic | 89.076 | 14 |

As of July 2019, the proposed model is ranked 9 out of 297 submissions. It achieves a high DSC and small distance-based metrics (95 HD and ABD) that are close to the leading methods. Distance based metrics assess the boundary delimitation and overall shape of the segmentation. In comparison, the DSC considers the spatial position of false negatives and false positives. Thus, the results show that the proposed method provides a good definition and contour of the segmented prostate.

Table 4. Quantitative results of the proposed method (2D-3D FCN) and top competing models on the PROMISE12 challenge test set.

| Model | DSC[%] | | | 95 HD [mm] | | | ABD [mm] | | | aRVD[%] | | |
|---|---|---|---|---|---|---|---|---|---|---|---|---|
| | Whole | Apex | Base | Whole | Apex | Base | Whole | Apex | Base | Whole | Apex | Base |
| 2D-3D FCN (ours) | 91.45 | 88.10 | 89.54 | 4.08 | 3.58 | 4.43 | 1.34 | 1.40 | 1.53 | 5.45 | 14.35 | 10.04 |
| HD_Net | 91.35 | 88.91 | 89.82 | 3.93 | 3.60 | 4.24 | 1.36 | 1.34 | 1.54 | 5.10 | 5.85 | 7.00 |
| Bowda-Net | 91.41 | 89.29 | 89.56 | 4.27 | 3.44 | 4.48 | 1.35 | 1.29 | 1.54 | 6.04 | 10.84 | 9.12 |
| Sunrise2014 | 90.58 | 87.89 | 89.65 | 4.95 | 3.90 | 4.10 | 1.59 | 1.47 | 1.48 | 5.81 | 9.79 | 5.94 |
| nnU-Net (I) | 91.61 | 88.05 | 90.29 | 4.00 | 3.79 | 4.05 | 1.31 | 1.46 | 1.45 | 7.13 | 14.97 | 8.29 |
| nnU-Net (II) | 91.56 | 88.52 | 89.59 | 4.17 | 3.77 | 4.42 | 1.30 | 1.39 | 1.49 | 6.93 | 13.61 | 10.17 |

The top performing models are deep learning architectures manually fitted for the prostate segmentation task (denoted as Design Type in Table 3). These architectures apply specialized mechanisms to identify the blurry boundaries between the prostate and surrounding structures. Differently from their approach, our ensemble has a simple and efficient structure that can be easily implemented, trained, and adapted to other datasets. Furthermore, the hyperparameters are set automatically without the laborious job of manual testing. The other two submissions are from an automatically-designed architecture, nnU-Net (I) and nnU-Net (II). The latter applies a ruled-based adaption for the preprocessing operations and hyperparameters of a group of U-Net architectures. The optimal network is an ensemble of a 2D and 3D U-Net architectures with $29.4 \times 10^6$ and $43.7 \times 10^6$ number of parameters, respectively. The first submission was trained with images from the challenge and an external dataset while the second was trained with images only from the challenge. Overall, our model performs better than the automatically-designed architectures and is considerably smaller ($1.6 \times 10^6$ parameters for the 2D FCN and $3.9 \times 10^6$ for the 3D FCN). Thus, showing that our model is better at automatically adapting to the prostate dataset and obtaining more efficient architectures.

## 4. CONCLUSIONS

In this paper, we presented a self-adaptive 2D-3D FCN ensemble for medical image segmentation that utilizes volumetric information and optimizes both the model's performance and size. The network consists of a 2D FCN that includes in-plane information and a 3D FCN that exploits volumetric information. The architecture of the FCN models are automatically constructed to maximize the segmentation accuracy and minimize the number of parameters in the model using a multiobjective evolutionary based algorithm. The proposed ensemble was tested in the task of prostate segmentation in the PROMISE12 Grand Challenge. Results demonstrate that our 2D-3D FCN ensemble achieved competitive outcomes in the test cases compared to manually-designed architectures while performing better than automatically-designed architectures. Therefore, our model successfully adapted its architecture and hyperparameters to the prostate dataset while demonstrating the potential of multiobjective evolutionary based algorithms for automatically designing smaller and efficient neural networks for medical image segmentation.

## ACKNOWLEDGMENTS

The authors would like to thank the Fulbright-Senescyt program for the support provided to Maria Baldeon-Calisto to pursue her Ph.D. degree.

## REFERENCES

[1] N. Sharma and L. M. Aggarwal, "Automated medical image segmentation techniques," *Journal of medical physics,* vol. 35, p. 3, 2010.

[2] J. Long, E. Shelhamer and T. Darrell, "Fully Convolutional Networks for Semantic Segmentation," in *IEEE Conference on Computer Vision and Pattern Recognition*, 2014.

[3] O. Ronneberg, P. Fischer and T. Brox, "U-Net: Convolutional Networks for Biomedical Image Segmentation," *arXiv:1505.04597,* 2015.

[4] V. Badrinarayanan, A. Kendall and R. Cipolla, "Segnet: A deep convolutional encoder-decoder architecture for image segmentation," *IEEE transactions on pattern analysis and machine intelligence,* pp. 2481-2495, 2017.

[5] O. Cicek, A. Abdulkadir, S. S. Lienkamp, T. Brox and O. Ronneberger, "3D U-Net: learning dense volumetric segmentation from sparse annotation," in *International conference on medical image computing and computer-assisted intervention*, 2016.

[6] K. Kamnitsas, C. Ledig, V. F. Newcombe, J. P. Simpson, A. D. Kane, D. K. Menon, D. Rueckert and B. Glocker, "Efficient multi-scale 3D CNN with fully connected CRF for accurate brain lesion segmentation," *Medical Image Analysis,* vol. 36, pp. 61-78, 2017.

[7] H. Chen, Q. Dou, L. Yu, J. Qin and P.-A. Heng, "VoxResNet: Deep voxelwise residual networks for brain segmentation from 3D MR images," *NeuroImage,* vol. 170, pp. 446-455, 2018.

[8] X. Li, H. Chen, X. Qi, Q. Dou, C.-W. Fu and P. A. Heng, "H-DenseUNet: Hybrid densely connected UNet for liver and liver tumor segmentation from CT volumes," *arXiv preprint arXiv:1709.07330,* 2017.

[9] P. Mlynarski, H. Delingette, A. Criminisi and N. Ayache, "3D convolutional neural networks for tumor segmentation using long-range 2D context," *Computerized Medical Imaging and Graphics,* pp. 60-72, 2019.

[10] T. Elsken, J. H. Metzen and F. Hutter, "Neural Architecture Search: A Survey," *Journal of Machine Learning Research,* pp. 1-21, 2019.

[11] B. Zoph and Q. V. Le, "Neural architecture search with reinforcement learning," *arXiv preprint arXiv:1611.01578,* 2016.

[12] H. Pham, M. Y. Guan, B. Zoph, Q. V. Le and J. Dean, "Efficient neural architecture search via parameter sharing," in *arXiv preprint arXiv:1802.03268*, 2018.

[13] B. Baker, O. Gupta, N. Naik and R. Raskar, "Designing neural network architectures using reinforcement learning," in *arXiv preprint arXiv:1611.02167*, 2016.

[14] R. Miikkulainen, J. Liang, E. Meyerson, A. Rawal, D. Fink, O. Francon, B. Raju, H. Shahrzad, A. Navruzyan and N. Duffy, "Evolving deep neural networks," *arXiv preprint arXiv:1703.00548,* 2017.

[15] E. Real, S. Moore, A. Selle, S. Saxena, Y. L. Suematsu, J. Tan, Q. V. Le and A. Kurakin, "Large-scale evolution of image classifiers," in *Proceedings of the 34th International Conference on Machine Learning*, 2017.

[16] M. Suganuma, S. Shirakawa and T. Nagao, "A Genetic Programming Approach to Designing Convolutional Neural Network Architectures," in *Proceedings of the Genetic and Evolutionary Computation Conference*, Berlin, 2017.

[17] K. Kandasamy, W. Neiswanger, J. Schneider, B. Poczos and E. P. Xing, "Neural architecture search with bayesian optimisation and optimal transport," in *Advances in Neural Information Processing Systems*, 2018.

[18] H. Liu, K. Simonyan and Y. Yang, "Darts: Differentiable architecture search," in *arXiv preprint arXiv:1806.09055*, 2018.

[19] S. Xie, H. Zheng, C. Liu and L. Lin, "SNAS: stochastic neural architecture," in *Proceedings of the International Conference on Learning Representations*, New Orleans, 2019.

[20] M. Wistuba, A. Rawat and T. Pedapati, "A Survey on Neural Architecture Search," in *arXiv preprint arXiv:1905.01392*, 2019.

[21] F. Isensee, J. Petersen, A. Klein, D. Zimmerer, P. F. Jaeger, S. Kohl, J. Wasserthal, G. Koehler, T. Norajitra, S. Wirkert and others, "nnu-net: Self-adapting framework for u-net-based medical image segmentation," *arXiv preprint arXiv:1809.10486,* 2018.

[22] A. Mortazi and U. Bagci, "Automatically designing CNN architectures for medical image segmentation," in *International Workshop on Machine Learning in Medical Imaging*, 2018.

[23] Y. Weng, T. Zhou, Y. Li and X. Qiu, "NAS-Unet: Neural Architecture Search for Medical Image Segmentation," *IEEE Access,* vol. 7, pp. 44247-44257, 2019.

[24] M. Baldeon and S. Lai-Yuen, "Adaresu-net: Multiobjective adaptive convolutional neural network for medical image segmentation," *Neurocomputing,* 2019.

[25] Z. Zhu, C. Liu, D. Yang, A. Yuille and D. Xu, "V-NAS: Neural Architecture Search for Volumetric Medical Image Segmentation," in *arXiv preprint arXiv:1906.02817*, 2019.

[26] S. Kim, I. Kim, S. Lim, W. Baek, C. Kim, H. Cho, B. Yoon and T. Kim, "Scalable Neural Architecture Search for 3D Medical Image Segmentation," in *arXiv preprint arXiv:1906.05956*, 2019.

[27] G. Litjens, R. Toth, W. van de Ven, C. Hoeks, S. Kerkstra, B. van Ginneken, G. Vincent, G. Guillard, N. Birbeck and J. Zhang, "Evaluation of prostate segmentation algorithms for MRI: the PROMISE12 challenge," *Medical Image Analysis,* vol. 18, pp. 359-373, 2014.

[28] S. Ioffe and C. Szegedy, "Batch normalization: Accelerating deep network training by reducing internal covariate shift," *arXiv preprint arXiv:1502.03167,* 2015.

[29] K. He, X. Zhang, S. Ren and J. Sun, "Identity mappings in deep residual networks.," in *European Conference on Computer Vision*, 2016.

[30] J. Tompson, R. Goroshin, A. Jain, Y. LeCun and C. Bregler, "Efficient object localization using convolutional networks," in *Proceedings of the IEEE Conference on Computer Vision and Pattern Recognition*, 2015.

[31] Q. Zhang and H. Li, "MOEA/D: A multiobjective Evolutionary Algorithm Based on Decomposition," *IEEE Transactions on Evolutionary Computations,* vol. 11, pp. 712-731, 2007.

[32] L. K. Hansen and P. Salamon, "Neural network ensembles," *IEEE Transactions on Pattern Analysis & Machine Intelligence,* pp. 993-1001, 1990.

[33] D. P. Kingma and J. Ba, "Adam: A method for stochastic optimization," *arXiv preprint arXiv:1412.6980,* 2014.

[34] F. Chollet, "Keras," 2015. [Online]. Available: https://github.com/keras-team/keras.